\documentclass[12pt]{article}
\usepackage{amsmath}
\usepackage{mathrsfs}
\usepackage{graphicx}
\usepackage{ amssymb }
\usepackage{multirow}
\usepackage[margin=2cm]{geometry} 
\usepackage{slashed}
\usepackage{caption}
\newcommand{\overbar}[1]{\mkern1.5mu\overline{\mkern-1.5mu#1\mkern-1.5mu}\mkern 1.5mu}

\def\abs#1{\left| #1\right|}            
\def\abs#1{\left| #1\right|}                    
\def\leftrightarrowfill{$\mathsurround=0pt \mathord\leftarrow \mkern-6mu
        \cleaders\hbox{$\mkern-2mu \mathord- \mkern-2mu$}\hfill
        \mkern-6mu \mathord\rightarrow$}
\def\dvec#1{\vbox{\ialign{##\crcr
        \leftrightarrowfill\crcr\noalign{\kern-1pt\nointerlineskip}
        $\hfil\displaystyle{#1}\hfil$\crcr}}}           

\def\VEV#1{\left\langle #1\right\rangle}        
\def\fracmm#1#2{{{#1}\over{#2}}}
  
\begin{document}

\thispagestyle{empty}

{\hbox to\hsize{
\vbox{\noindent January 2018 \hfill IPMU17-0141 }}
\noindent revised version \hfill }

\noindent
\vskip2.0cm
\begin{center}

{\Large\bf Removing instability of inflation in Polonyi-Starobinsky 
\vglue.1in
supergravity by adding FI term
}

\vglue.3in

Yermek Aldabergenov~${}^{a}$  and Sergei V. Ketov~${}^{a,b,c}$ 
\vglue.1in

${}^a$~Department of Physics, Tokyo Metropolitan University, \\
Minami-ohsawa 1-1, Hachioji-shi, Tokyo 192-0397, Japan \\
${}^b$~Institute of Physics and Technology, Tomsk Polytechnic University,\\
30 Lenin Ave., Tomsk 634050, Russian Federation \\
${}^c$~Kavli Institute for the Physics and Mathematics of the Universe (IPMU),
\\The University of Tokyo, Chiba 277-8568, Japan \\
\vglue.1in
aldabergenov-yermek@ed.tmu.ac.jp, ketov@tmu.ac.jp
\end{center}

\vglue.3in

\begin{center}
{\Large\bf Abstract}
\end{center}
\vglue.1in
\noindent  Polonyi-Starobinsky (PS) supergravity is the $N=1$ supergravity model of Starobinsky inflation with spontaneous supersymmetry breaking (after inflation) due to Polonyi superfield, and inflaton belonging to a massive vector supermultiplet. The PS model is used for an explicit realization of the (super-heavy) gravitino dark matter scenario in cosmology. We find a potential instability in this model, and offer a mechanism for its removal by adding a Fayet-Iliopoulos (FI) term.

\newpage

\section{Introduction}

The importance of the inflationary model building in supergravity stems from the natural objective to unify gravity with particle physics beyond the Standard Model of elementary particles and beyond the Standard ($\Lambda$CDM) Model of cosmology. There are two different approaches to the inflationary model building in supergravity, depending upon assignment of inflaton to either a (massive) chiral multiplet or to a  (massive) vector mutiplet. As regards the first (and mostly studied) approach, see e.g., \cite{Yamaguchi:2011kg,Ketov:2012yz} for a review. The second (less studied) approach is based on the observation that a massive $N=1$ vector multiplet has only one real scalar, so that there is no need for stabilization of another (non-inflaton) scalar, contrary to the first approach (it is known as the $\eta$-problem). The minimal supergravity models with inflaton belonging to a massive vector multiplet were proposed in Refs.~\cite{Farakos:2013cqa,Ferrara:2013rsa} by using the non-minimal self-coupling of a vector multiplet, which is paramaterized by an arbitrary real function \cite{VanProeyen:1979ks}.  These models can accommodate any desired values of the Cosmic Microwave Background (CMB) observables ($n_s$ and $r$), because the corresponding single-field (inflaton) scalar potential is given by the derivative squared of that arbitrary real function. However, all models \cite{Farakos:2013cqa,Ferrara:2013rsa} have the vanishing vacuum energy after inflation, i.e. the vanishing cosmological constant, and the vanishing Vacuum Expectation Value (VEV) of the auxiliary fields, so that supersymmetry (SUSY) is restored after inflation and only a Minkowski vacuum is allowed. These models were improved in \cite{Aldabergenov:2016dcu,Aldabergenov:2017bjt}, where an extra (Polonyi) chiral superfield with a linear superpotential was added, and it was demonstrated to lead to a spontaneous SUSY breaking and a de-Sitter vacuum after inflation. The particular improved model accommodates the Starobinsky inflationary potential, so it is called the {\it Polonyi-Starobinsky} (PS) supergravity. The latter was  employed in \cite{Addazi:2017ulg} for a construction of a viable scenario of the super-heavy gravitino dark matter with the spontaneously broken high-scale SUSY and R-symmetry.

In this letter we observe a dangerous instability in the scalar potential of the PS supergravity, and offer a cure for its removal by adding a Fayet-Iliopoulos (FI) term \cite{Fayet:1974jb} together with its SUSY- and gauge-invariant completion  \cite{Cribiori:2017laj}
without gauged R-symmetry.

\section{PS model}

The Lagrangian of PS supergravity in a curved $N=1$ superspace with a K\"ahler potential $K$, a superpotential $\mathcal{W}$,  and a real  function $J=J(V)$ of the vector superfield $V$, is given by \cite{Aldabergenov:2016dcu}~\footnote{Our notation is standard, as in  Ref.~\cite{Wess:1992cp}, with the spacetime signature $(-,+,+,+)$.  The reduced Planck mass $M_P$ and the gauge coupling $g$ of the vector multiplet are both set to $1$ for simplicity.} 
\begin{equation} \label{lag}
\mathcal{L}=\int d^2\Theta 2\mathcal{E}\left[\frac{3}{8}(\bar{\mathcal{D}}^2-8\mathcal{R})e^{-\frac{1}{3}(K+2J)}+\frac{1}{4}W^\alpha W_\alpha+\mathcal{W}\right]+\text{h.c.}~,
\end{equation}
where $2\mathcal{E}$ is the chiral density superfield, $\mathcal{R}$ is the chiral scalar curvature superfield, and $W_\alpha\equiv-\frac{1}{4}(\overbar{\mathcal{D}}\overbar{\mathcal{D}}-8\mathcal{R})\mathcal{D}_\alpha V$ is the chiral vector superfield strength. 

Its Polonyi part is defined by the following K\"ahler potential and the superpotential \cite{Polonyi:1977pj}:
\begin{equation} \label{polonyi}
K= \Phi\overbar{\Phi}~,\qquad \mathcal{W}=\mu(\Phi +\beta)~,
\end{equation}
where $\Phi$ is the Polonyi superfield whose lowest component $A$ is a complex scalar, $\mu$ is the (free) parameter fixing the scale
of SUSY breaking, while the parameter $\beta$ is determined by the SUSY breaking vacuum solution (see below).

The bosonic part of the Lagrangian (\ref{lag}) reads \cite{Aldabergenov:2016dcu,Aldabergenov:2017bjt}
\begin{equation} \label{complag}
e^{-1}\mathcal{L}=-\frac{1}{2}R-\partial_m A\partial^m\bar{A}-\frac{1}{4}F_{mn}F^{mn}-\frac{1}{2}J''\partial_mC\partial^mC-\frac{1}{2}J''B_mB^m-\mathcal{V},
\end{equation}
with the scalar potential
\begin{equation} \label{Vpot}
\mathcal{V}=\frac{1}{2}J'^2+\mu^2 e^{K+2J}\left\{ |\bar{A}A+A\beta+1|^2-\left(3-2\frac{J'^2}{J''}\right)|A+\beta|^2\right\}~,
\end{equation}
where $C$ is the (real) scalar component of the real superfield $V$, $B_m$ is its vector component having the (abelian) field strength $F_{mn}$, and the primes denote the derivatives with respect to $C$. 

In order to obtain the (D-type) Starobinsky inflationary potential and the canonical kinetic term of the inflaton field, one can choose the function $J(C)$ and redefine $C$ as \cite{Ferrara:2013rsa}
\begin{equation}
C=-\exp\left(\sqrt{2/3}\phi\right)~,~~~J=-\frac{3}{2}[\log(-C)+C]~,~~~J'=-\frac{3}{2}\left(C^{-1}+1\right)~,~~~J''=\frac{3}{2}C^{-2}~,\label{CJdef}
\end{equation}
in terms of the canonical scalar $\phi$ playing the role of inflaton. Then the full scalar potential in 
PS supergravity is a sum of the $D$-type and $F$-type terms as follows:
\begin{gather}
V_D=\frac{9}{8}g^2 \left(1-e^{-\sqrt{2/3}\phi}\right)^2~, \nonumber\\
V_F=\mu^2 \exp\left(\bar{A}A-\sqrt{6}\phi+3e^{\sqrt{2/3}\phi}\right)\times \label{VFStarob}\\
\times \left\{ \bar{A}A+A\beta+1|^2-3\left[\left(1-e^{\sqrt{8/3}\phi}\right) \left(1-e^{-\sqrt{2/3}\phi}\right)^2\right]|A+\beta|^2\right\}~,\nonumber
\end{gather}
where $V_D$ is the celebrated (Starobinsky) potential responsible for (large-field, slow-roll) inflation. We have restored above the
gauge coupling constant $g$ that is proportional to the inflaton mass (in turn, the latter is fixed by CMB observations to the order
of $10^{-5}M_P$). The coupling constant $\mu$ is arbitrary and determines the scale of spontaneous SUSY breaking,
as well as the masses of Polonyi and gravitino fields in our model \cite{Aldabergenov:2016dcu}.~\footnote{As regards the phenomenological applications of Polonyi-Starobinsky supergravity, which are based on the model under consideration, 
see Ref.~\cite{Addazi:2017ulg}.}

The Starobinsky inflation driven by the D-term above can, however, be affected (and even destroyed) by the
F-type term, unless the latter is suppressed during inflation, 
\begin{equation} \label{supp}
|V_F| \ll |V_D|~~,
\end{equation}
because the inflaton and Polonyi fields are mixed in the $V_F$ of Eq.~(\ref{VFStarob}). The simplest way of getting the condition
(\ref{supp}) is to assume $\mu \ll g$. However, on the physical side, this would imply the masses of gravitino and Polonyi to be much less than the inflaton mass. In turn, it leads to the well known overproduction problems with gravitino and Polonyi particles that overclose the universe. The super-heavy gravitino dark matter scenario \cite{Addazi:2017ulg}, based on the model under investigation, avoids both Polonyi and gravitino problems by considering a High-scale SUSY breaking that implies a large value
of $\mu$ comparable with the value of $g$. Hence, the condition (\ref{supp}) should be enforced in another way.

Given $\mu\sim g$, there are two dangerous terms in Eq.~(\ref{VFStarob}): first, there is the factor 
$ \exp\left(3e^{\sqrt{2/3}\phi}\right)$ growing very fast with $\phi$, and second, there is another fast-growing factor 
$e^{\sqrt{2/3}\phi}$  in front of the 2nd term in the curved brackets, which disturbs the vacuum condition on the Polonyi field (the curved bracket vanishes when $A=\VEV{A}$ in the absence of the $J$-dependent factor inside the curved brackets of Eq.~(\ref{Vpot}) --- see Ref.~\cite{Aldabergenov:2016dcu} for details). This claim is supported by numerical calculations of the initial 
value of $\phi$-field during Starobinsky inflation --- see e.g., Ref.~\cite{Ellis:2015pla}. We find that the initial (maximal) value of 
$\phi$ for the e-foldings number $N_e=50$ is about 5.16, while the corresponding factor $e^{\sqrt{2/3}\phi}$ is approximately equal 
to $\frac{4}{3}N_e +1.1$, i.e. can be as high as 67.7.

We describe this situation as an "instability" because the unsuppressed F-term may result in a considerable deviation of the inflationary trajectory (of inflaton $\phi$) from its desired (Starobinsky) solution, as well as breaking down the slow-roll conditions. Therefore, (i) the Polonyi field should be strongly stabilized in its vacuum, and (ii) both dangerous (fast-growing and $J$-dependent) factors should be removed. In order to achieve the goal (i), we assume a large Polonyi mass (beyond the Hubble value). However, as regards a cure to the remaining problem (ii), we need an additional resource beyond the original model  \cite{Aldabergenov:2016dcu}. In the next Section we propose to employ a FI term for that  purpose.~\footnote{The alternative may be a modification of the K\"ahler potential of the Polonyi superfield  and adding its non-minimal coupling to the inflaton superfield, e.g., along the lines of Ref.~\cite{Nakayama:2016eq}.}

\vglue.2in

\section{Improved PS model with FI term}

We are aware of the fact that the standard FI term \cite{Fayet:1974jb} in supergravity does not transform to a total derivative under local SUSY (unlike rigid SUSY), so that it requires a SUSY completion. Such a completion was computed the long time ago \cite{Freedman:1976uk} by using Noether (trial-and-error) procedure, which amounts to adding the gravitino-photino mixing term and the vector gauge connection in the gravitino supercovariant derivative. The latter amounts to the gauging of R-symmetry, so that it requires the vanishing gravitino mass and, hence, is obviously inconsistent with our approach.~\footnote{See also 
Ref.~\cite{Binetruy:2004hh}, as regards other difficulties of exploiting the FI term  \cite{Fayet:1974jb} in cosmology and string theory.}

However, the FI completion \cite{Freedman:1976uk} is not unique, and there exist {\it another} linearly-realized SUSY completion 
\cite{Cribiori:2017laj} of a FI term, without gauging the R-symmetry and allowing for a non-vanishing gravitino mass, which is perfectly suitable for our purposes! The extra (FI) term to be added to our Lagrangian (\ref{lag}) reads \cite{Cribiori:2017laj} 
\begin{equation} \label{lagFI}
\mathcal{L}_{\text{FI}}=8\xi\int d^4\theta E\frac{W^2\overbar{W}^2}{\mathcal{D}^2W^2\overbar{\mathcal{D}}^2\overbar{W}^2}\mathcal{D}^\alpha W_\alpha
\end{equation}
with an constant (real) parameter $\xi$. This term is {\it manifestly\/} SUSY- and gauge-invariant, does {\it not\/}  include higher spacetime derivatives of the field components, but has the inverse powers of the auxiliary field $D$ (up to the forth order) in the
{\it fermionic} sector only. We set all fermions to be zero in our discussion, so that the scalar $D$ enters the bosonic action as a quadratic polynomial. The K\"ahler gauge invariance, also broken by the FI term above, can be restored by further modifications of the K\"ahler gauge transformations \cite{Cribiori:2017laj}.

Our idea, in order to suppress the dangerous terms in $V_F$ causing the instability, is to modify the function $J$ in the PS supergravity with the FI term, and simultaneously compensate the resulting change in $V_D$, in order the keep the Starobinsky potential $V_D$ (in the unitary gauge $H=1$). With an arbitrary $J$-function and the FI term (\ref{lagFI}), the scalar potential  (\ref{VFStarob}) gets modified as
\begin{gather}
V_D=\frac{g^2}{2}\left[ J'+\xi e^{\frac{1}{3}(K+2J)}\right]^2~,\label{VD}\\
V_F=\mu^2 e^{\bar{A}A+2J}\left\{ |\bar{A}A+A\beta+1|^2-\left(3-2\frac{J'^2}{J''}\right)|A+\beta|^2\right\}~.\label{VF}
\end{gather}

Demanding the $V_D$ to reproduce the Starobinsky potential yields the first-order non-linear differential equation as
\begin{equation} \label{diffe}
\fracmm{dJ}{dC}+\xi e^{\frac{1}{3}(K+2J)} = - \fracmm{3}{2} \left( 1 + \fracmm{1}{C} \right) ~.
\end{equation}
Since we want Polonyi field $A$ to stay in its minimum at $A=\VEV{A}$,  we can introduce the effective (field-dependent) FI term
$\tilde{\xi}(A,\bar{A}) =\xi e^{\frac{1}{3}K(A,\bar{A})}$ together with its VEV, $\tilde{\xi}=\xi e^{\frac{1}{3}K(\VEV{A},
\VEV{\bar{A}})}$, and rewrite (\ref{diffe}) to the form of a single equation on the effective $J$-function as
\begin{equation} \label{diffe2}
\fracmm{dJ}{dC}+\tilde{\xi} e^{\frac{2}{3}J} = - \fracmm{3}{2} \left( 1 + \fracmm{1}{C} \right) ~.
\end{equation}
The case without the FI term, considered above, is reproduced when $\tilde{\xi}=0$, which leads to the asymptotic behaviour 
$J\sim -\frac{3}{2}C >0$ for large negative $C$, that is exactly the cause of instability. 

This instability will be removed when the function $J$ would approach a constant instead, because large negative values of $C$ exactly correspond to a plateau (slow-roll) of Starobinsky inflation, according to (\ref{CJdef}) and (\ref{VFStarob}). Indeed,  Eq.~(\ref{diffe2}) can be easily integrated at $\abs{C^{-1}}\ll 1$, with the result
\begin{equation} \label{dfres}
J(C) \approx J_{\infty} - \frac{3}{2} \ln \left( 1 - e^{C-C_0}\right) ~,
\end{equation}
where $C_0$ is the integration constant, and we have used the notation
\begin{equation} \label{not1}
J_{\infty}=\frac{3}{2} \ln \left(\fracmm{3}{-2\tilde{\xi} }\right)~.
\end{equation}
As is clear from (\ref{dfres}) and (\ref{not1}), demanding 
 \begin{equation} \label{sign}
\xi < 0
\end{equation}
implies $\tilde{\xi} < 0$ also, whereas the function $J$ fast approaches the constant $J_{\infty}$ from above, with $C$ taking large negative values. It is worth noticing that $J_{\infty}=0$ at the "critical" value $\tilde{\xi}=-3/2$.

In the case of demanding the $V_D$ to have the form of the Starobinsky potential as above, the VEV of the $D$ field 
in the Minkowski vacuum vanishes, so that there may be a problem with consistency of the {\it fermionic} terms with the negative powers of $D$, as was already noticed in \cite{Cribiori:2017laj}. We would like to mention here that this problem can be easily cured by a small uplifting of the Minkowski vacuum to a de Sitter vacuum (after inflation) via a slight modification of the function on the r.h.s. of Eq.~(\ref{diffe}), leading to a (small) positive cosmological constant. Then the VEV of $D$ will be non-vanishing and the fermionic terms will be well-defined.

The stability analysis of the scalar potential $V_F$ was already performed in Ref.~\cite{Aldabergenov:2016dcu} in the case of
$\xi=0$, and it does not significantly change with the FI term. It is, however, instructive to check how the condition
\begin{equation} \label{smr}
\fracmm{(J')^2}{J''} \ll 1
\end{equation}
is satisfied in the case above.  We find for large negative values of $C$ that
 \begin{equation} \label{smr2}
\fracmm{(J')^2}{J''} \approx -\fracmm{3}{2}C^{-1}
\end{equation}
{\it independently} upon the value of $\tilde{\xi}$, so that the condition (\ref{smr}) is always satisfied for large $\abs{C}\gg 1$.

Our (FI-modified) inflationary scalar potential of PS  supergravity during slow-roll reads
\begin{gather}
\mathcal{V}=\frac{9}{8}g^2M_P^4\left(1-e^{-\sqrt{2/3}\phi/M_P}\right)^2+\mu^2M_P^{-2}\exp\left(M_P^{-2}\bar{A}A
+2J_{\infty} \right) \times \nonumber \\
\times  \left\{  |\bar{A}A+A\beta+M_P^{2}|^2 - 3M_P^{2}\left(1-e^{-\sqrt{2/3}\phi/M_P}\right)  |A+\beta|^2
\right\} \equiv V_D + V_F~~,
\end{gather}
where we have restored the dependence upon the reduced Planck mass $M_P$. At  large values of $\phi$ (and fixed $\bar{A},A$), the $V_F$ goes to zero, while $V_D\rightarrow 9g^2M_P^4/8$.

\section{Conclusion}

Our new results are given by Sec.~3, their motivation is in Sec.~1, and their physical significance is explained in the Abstract. It was achieved by introducing the FI term and changing the $J$-function under the condition $\xi <0$. After eliminating the auxiliary fields,  taking the limit of $\xi\to 0$ is impossible. It is expected that our full action can be rewritten in terms of constrained superfields in the
context of nonlinear realizations of local SUSY \cite{Ivanov:1978mx,Komargodski:2009rz}, similarly to the cases studied in \cite{Cribiori:2017laj}.

Our PS supergravity model can be part of a more general (and more realistic) theory including more matter and the hidden sector to
be suitable for phenomenological applications. The gauge-invariant formulation of our model \cite{Aldabergenov:2017bjt} is suitable for further generalizations, including unification of inflation with the supersymmetric Grand Unified Theories (GUTs) in the context of supergravity, when the GUT gauge group has an  abelian $U(1)$ factor (favoured by superstring compactifications), by including a positive definite inflaton scalar potential, a spontaneous SUSY breaking and a de Sitter vacuum after inflation. Our approach does not preserve the R-symmetry. In particular, it favours the super-GUTs without monopoles \cite{Ketov:1996bm}.  Details of the relation of our model to super-GUTs and reheating are dependent upon the way how the fields present in our model interact with the super-GUT fields, while all that is highly model-dependent.   Our models can be further extended in the gauge-sector to the Born-Infeld-type gauge theory coupled to supergravity and other matter, along the lines of Ref.~\cite{Abe:2015fha}, thus providing further support towards their possible origin in superstring (flux) compactification. As regards a possible {\it dynamical} origin of FI term, see e.g., Ref.~\cite{Domcke:2017rzu}.

\section*{Acknowledgements}

Y.A. was supported by a scholarship from the Ministry of Education, Culture, Sports, Science and Technology (MEXT) in Japan. S.V.K. was supported by a Grant-in-Aid of the Japanese Society for Promotion of Science (JSPS) under No.~26400252,  the Competitiveness Enhancement Program of Tomsk Polytechnic University in Russia, and the World Premier International Research Center Initiative (WPI Initiative), MEXT, Japan. The authors thank Hiroyuki Abe and  Alexandros Kehagias for discussions and correspondence, and the anonymous referee for his critical remarks.

\vglue.2in

\bibliographystyle{utphys} 

\providecommand{\href}[2]{#2}\begingroup\raggedright
\endgroup

\end{document}